# Decomposition scheme matters more than you may think

Anna NASZODI[1]

This paper promotes the application of a path-independent decomposition scheme. Besides presenting some theoretical arguments supporting this decomposition scheme, this study also illustrates the difference between the path-independent decomposition scheme and a popular sequential decomposition with an empirical application of the two schemes. The empirical application is about identifying a directly unobservable phenomenon, i.e. the changing social gap between people from different educational strata, through its effect on marriages and cohabitations. It exploits census data from four waves between 1977 and 2011 about the American, French, Hungarian, Portuguese, and Romanian societies. For some societies and periods, the outcome of the decomposition is found to be highly sensitive to the choice of the decomposition scheme. These examples illustrate the point that a careful selection of the decomposition scheme is crucial for adequately documenting the dynamics of unobservable factors.

Keywords: counterfactual decompositions; Oaxaca-Blinder decomposition scheme; path-independent decomposition scheme; social stratification

**Introduction**

Decompositions are widely applied by researchers who want to quantify the relative strength of different contributing factors to a given outcome variable. Often, decomposition exercises are performed with another aim, i.e., to study a phenomenon that is not observable directly. The most famous example is provided by the seminal papers by Blinder (1973), and Oaxaca (1973), who study discrimination through its

---

[1] Anna Naszodi, European Commission's Joint Research Centre (Address: Via Fermi 2749, I-21027, Ispra, Italy) and member of the Centre for Economic and Regional Studies (KRTK) (Address: Toth Kalman utca 4, H-1097, Budapest, Hungary).
Emails: anna.naszodi@ec.europa.eu, anna.naszodi@gmail.com.
Phone: +39-0332783948. Fax: +39-0332785733.

The views expressed in this paper are those of the author and do not necessarily reflect the official views of the European Commission.

effect on observable wages. They do it by decomposing wage gap between people belonging to different groups (Black/White or male/female) into its observable determinants and an unobservable determinant, i.e., discrimination against a certain group.

Oaxaca and Blinder developed a decomposition technique for this purpose in 1973. Since then it has been applied not only to several other topics, but also in other disciplines (see O'Donnell et al., 2008). It enjoys an almost unbroken popularity.

However, some scholars, typically those working in the field of labor economics, have already started to use an alternative decomposition scheme. This alternative scheme is inspired by the original Oaxaca-Blinder decomposition scheme: it is a modified Oaxaca-Blinder decomposition scheme that is free of certain limitations of the original scheme. Although the limitations in question were already pointed out in the late 2000s/early 2010s (see e.g. Jann 2008, Elder, Goddeeris, and Haider 2010, and Biewen 2012), many social scientists routinely apply the original Oaxaca-Blinder decomposition scheme even nowadays.[2]

*The purpose of this paper is to promote the modified Oaxaca-Blinder decomposition scheme by comparing it with the original Oaxaca-Blinder decomposition scheme.*

First, I present the arguments by Biewen (2012) put forth in favour of the modified decomposition scheme. His arguments are not specific to any application of the decomposition techniques since those are stated in an abstract mode.

Then, I present a new application of the methods that illustrates well two points. One is that some qualitative findings obtained by the modified decomposition scheme are robust to changes in the sample. The other is that the same findings are sensitive to the choice of the decomposition scheme. This new application is not taken from the well explored discrimination literature, but from the literature on assortative mating.[3]

---

[2] For instance, Google Scholar gives more than 500 results when searching for the "Oaxaca-Blinder decomposition" phrase in studies published in year 2019.

**Two decomposition formulas**

This section presents two decomposition schemes under the simplest dynamic setup. At time *0* and *1* the value of an observable outcome variable is determined by function *f* and two factors *x*, and *y*.

The decomposition problem can be stated as follows. We would like to know, what share of the change in function *f(x, y)* is due to changes in *x*, and *y*.

The *original Oaxaca-Blinder decomposition scheme* is either

$$f(x_1,y_1) - f(x_0,y_0) = \overbrace{[f(x_1,y_0) - f(x_0,y_0)]}^{\text{due to } \Delta x} + \overbrace{[f(x_1,y_1) - f(x_1,y_0)]}^{\text{due to } \Delta y}, \text{ or} \quad (1)$$

$$f(x_1,y_1) - f(x_0,y_0) = \overbrace{[f(x_1,y_1) - f(x_0,y_1)]}^{\text{due to } \Delta x} + \overbrace{[f(x_0,y_1) - f(x_0,y_0)]}^{\text{due to } \Delta y}, \quad (2)$$

depending on the assumed sequence of the changes in *x* and *y*. In particular, Eq. (1) corresponds to the case, where *x* is assumed to change prior to *y*. And Eq. (2) corresponds to the case, where *y* is assumed to change first.

---

[3] In this strand of literature, Dupuy and Weber (2018) and Eika et al (2019) used the original Oaxaca-Blinder decomposition scheme recently to quantify the effect of assortative mating on income inequality across households.

The formula for the *modified scheme* is

$$f(x_1,y_1) - f(x_0,y_0) = \overbrace{[f(x_1,y_0) - f(x_0,y_0)]}^{\text{due to } \Delta x} + \overbrace{[f(x_0,y_1) - f(x_0,y_0)]}^{\text{due to } \Delta y} +$$

$$+ \underbrace{[f(x_1,y_1) - f(x_1,y_0) - f(x_0,y_1) + f(x_0,y_0)]}_{\text{due to the joint effect of } \Delta x \text{ and } \Delta y}, \quad (3)$$

By comparing formula (3) with (1) and (2), one can see that the original Oaxaca-Blinder decomposition scheme attributes the joint effect of $\Delta x$ and $\Delta y$ in (3) either to the first factor, or to the second factor.

If the interaction effect $[f(x_1,y_1) - f(x_1,y_0) - f(x_0,y_1) + f(x_0,y_0)]$ is zero, then the two decomposition schemes provide the same pair of components. However, if the interaction effect is not negligible, the application of the two decomposition schemes can result in quite different pair of components as it will be illustrated in the empirical part of this paper.

**Theoretical considerations for selecting the decomposition scheme**

Biewen (2012) has two main arguments supporting the use of the decomposition scheme of (3) instead of (1) or (2), in case the interaction effect is not negligible.

First, unlike the components estimated by the original Oaxaca-Blinder decomposition scheme, the components obtained by the modified scheme do not depend on any assumption about the sequence of changes in the factors. In other words, the latter is a path-independent decomposition scheme. Obviously, the use of this scheme is preferred whenever we do not know the sequence of changes in the factors.

Second, the modified decomposition scheme does not attribute the interaction effect to any of the parts capturing the ceteris paribus effects of the factors. By contrast, the original Oaxaca-Blinder decomposition scheme does so mechanically despite the fact that the interaction effect represents a joint, inseparable effect.

For instance, a cup of instant coffee can be prepared neither without the coffee powder, nor without water. Still, by using the original Oaxaca-Blinder decomposition scheme, we would have to attribute the pleasure of drinking coffee either to the water component or the coffee powder component, depending on whether we add water to a spoon of instant coffee or we add a spoon of instant coffee to the hot water.

**An empirical application**

The decomposition problem examined empirically in this section is as follows. The share of couples where the spouses have the same education level is changing over time. It is driven (i) partly by an *unobservable factor, the changing mating preferences of young adults in a society*,[4] (ii) partly by an observable factor, the changing structural availability of men and women with given educational attainment, and (iii) partly by the joint effect of availability and preferences (see e.g. Kalmijn, 1998).

Naszodi and Mendonca (2021) developed a method for decomposing the effects of each of these drivers on the outcome variable in order to identify the unobservable factor.[5] To illustrate the application of their method, they performed a decomposition analysis for the U.S. for the period between 1980 and 2010 by using decennial census data.

They find that young American adults from the early Baby Boomers were more picky regarding the education level of their spouses than the members of the late Baby Boomers were at the same age. Another finding of Naszodi and Mendonca is that young

---

[4] By changing preference in a society, it is meant the following. The young adults from a generation with given preferences are replaced over time in the marriage market by the new young adults from a younger generation with different preferences.

[5] In the paper by Naszodi and Mendonca (2021), marital preferences are assumed to be characterized by the generalized Liu-Lu measure calculated from the contingency table of couples. If the assorted trait is a dichotomous variable, then the generalized Liu-Lu measure is equivalent to the original scalar-valued Liu-Lu measure proposed by Liu and Lu (2006). However, if the assorted trait is an ordered polytomous variable, the generalized Liu-Lu measure is a matrix-valued, multidimensional measure (see Naszodi and Mendonca 2021).

American adults from the late GenerationX had much more intense "desire" for having a well-educated spouse than the early GenerationX had when aged the same. [6]

In this paper, I replicate the empirical analysis of Naszodi and Mendonca (2021) using different samples. *My primary purpose with the replications is to investigate the sensitivity of their findings to the choice of the decomposition scheme.* In addition, I check the robustness of the results to the choice of the sample itself, and the combination of sample selection and choice of the decomposition scheme.

Following Naszodi and Mendonca (2021), I use decennial census data from four waves on the educational attainment of married couples and cohabiting couples from the Integrated Public Use Microdata Series (IPUMS). The couples are selected to my samples with the same criterion used by Naszodi and Mendonca (2021): the analyzed young couples are those, where the male partners are aged between 30 and 34. However, the data I use are not only about the U.S., but those also cover couples in four European countries, France, Hungary, Portugal, and Romania. [7]

Figure 1a shows the outcome of the decompositions for each decades and each country separately. It is worth noting that all of its five black bars belonging to the 1980s are in the negative range. Moreover, all the black bars belonging to the 2000s are in the positive range and all are relatively high. These results suggest that the qualitative empirical findings of Naszodi and Mendonca (2021) about the differences in the "desire for marry your like" across the generations examined are not sensitive to the choice of the country analyzed. [8]

---

[6] In the analysis of Naszodi and Mendonca (2021), the early Baby Boomers are represented by those who were born between 1946 and 1950; the late Baby Boomers are represented by those who were born between 1956 and 1960. Finally, the individuals in the samples of early GenerationX and the late GenerationX were born between 1966 and 1970; and 1976 and 1980, respectively.

[7] For other European countries, I found no data of good quality on the educational distribution of couples in IPUMS.

[8] Sociologists would interpret these findings as follows (see Lichter and Qian 2019). The integrity of the marriage markets/mating markets has increased universally over a certain

Let us turn to checking the *robustness of the decompositions with respect to the decomposition scheme*. There is one country for which the outcome of the decomposition is highly sensitive to the choice of the decomposition scheme. This country is Romania. The lack of robustness can be best shown by comparing its changing preference effect with its interaction effect, both exerted over the entire three decades analyzed. In case of Romania, these two effects are of the same magnitude, but they have opposite signs (see the striped bar and the black bar in Figure 1b belonging to Romania).

Accordingly, the path-independent decomposition scheme estimates the changing preferences to have had a substantial negative ceteris paribus effect on the share of educationally homogamous Romanian couples over the entire three decades. By contrast, the original Oaxaca-Blinder decomposition scheme suggests that this effect is negligible when it arbitrarily attributes the interaction effect to the changing preferences.

By looking at the results of the decade-specific, short-horizon decompositions, we find some other examples for the lack of robustness with respect to the decomposition scheme (see Figure 1a). For instance, whether the interaction effect between 1999 and 2011 is attributed to the changing educational distributions or not, does not seem to be neutral to the sign of the latter effect in case of France. Similarly, it matters also for Hungary for the period between 1990 and 2001. See the striped bars and the white bars in Fig.1a belonging to France (1999-2011) and Hungary (1990-2001).

The most illustrative example is again with Romania: if we attributed its interaction effect between 1977 and 1992 to the changing preferences by using the original Oaxaca-Blinder decomposition scheme, then the sign of the latter effect would change from negative to positive (see Figure 1a). Therefore, the finding about the universally weakening intensity of the "desire for marry your like" among the young

---

period. This period was followed by another period characterized by the common trend of increasing segmentation of these markets along the educational dimension. If we also add that ones' education level is a proxy for ones' ability to generate income, then these findings can be related not only to the dynamics of the social gap between different educational groups, but also to the dynamics in income inequality.

adults in the 1980s would be obscured by the application of the original Oaxaca-Blinder decomposition scheme.

To sum up, the example of Romania between 1977 and 2011 is a strong case for the sensitivity of the estimated components to the choice of the decomposition scheme. Moreover, it is not the only case; there are other examples as well (e.g. the examples with France and Hungary). Interestingly, one would not find it important to choose between the decomposition schemes in the case when the empirical analysis is performed only on the American data and/or the Portuguese data.

Finally, let us check the *robustness of the results with respect to the age of the couples*. In particular, we replicate the decompositions on a sample of couples, where the male partners are aged between 25 and 39 years (as opposed to being in the 30-34 years age category). The motivation for performing this robustness check is that the age category is broad in other papers in the empirical literature (see Lichter & Qian 2019; Eika et al. 2019).

The results of this robustness check are presented by Figure 2. The outcomes of the decompositions are qualitatively the same as those shown by Figure 1. Most importantly, those findings that have proved to be sensitive to the choice of the decomposition scheme with the sample of 30-34 years old male partners remained sensitive after changing the age group examined (see Figure 2a for France between 1999 and 2011, Hungary between 1990 and 2001, Romania between 1977 and 1992; and Figure 2b for Romania between 1977 and 2011).

**Conclusion**

This paper compared a path-independent decomposition scheme with the commonly applied decomposition scheme developed by Oaxaca and Blinder. First, it presented some theoretical arguments put forth by Biewen (2012) supporting the application of the path-independent decomposition scheme as opposed to the original Oaxaca-Blinder scheme.

Second, this paper compared the two schemes in their empirical applications to a specific decomposition problem. The empirical application illustrates the following point: while a key qualitative finding obtained by the path-independent scheme can be robust to changes in the sample, the same might not hold true when using the original Oaxaca-Blinder scheme.

In conclusion, there are not only theoretical reasons for applying the path-independent decomposition scheme, but its robustness to certain changes of the sample makes it appealing also from an empirical perspective. Biewen (2012), Elder et al. (2010), and Jann (2008) illustrate the difference across various decomposition schemes by analyzing discrimination. They also find noticeable differences in the results obtained with the schemes compared. Their finding, together with the finding of this paper, suggests that the lack of robustness to the decomposition scheme is not specific either to discrimination analyses or to the assortative mating analyses. Therefore, the careful selection of the decomposition scheme can be important from the perspective of other applications as well.

Oaxaca, R. (1973). Male-female wage differentials in urban labor markets. *International Economic Review, 14 (3)*, 693-709.

O'Donnell, O., Doorslaer, E., Wagstaff, A., & Lindelow, M. (2008). Explaining differences between groups: Oaxaca decomposition. *Analyzing health equity using household survey data: A guide to techniques and their implementation*, 147-157.

Figure 1. Decomposition of changes in the share of educationally homogamous couples between the late 1970s/early 1980s and the early 2010s - male partners are aged 30-34 years.

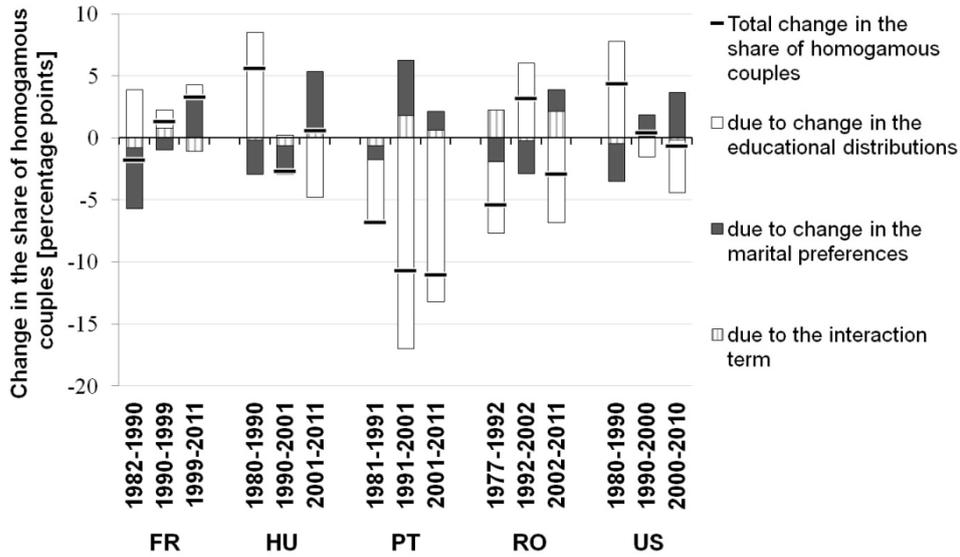

a) Short-horizon decompositions

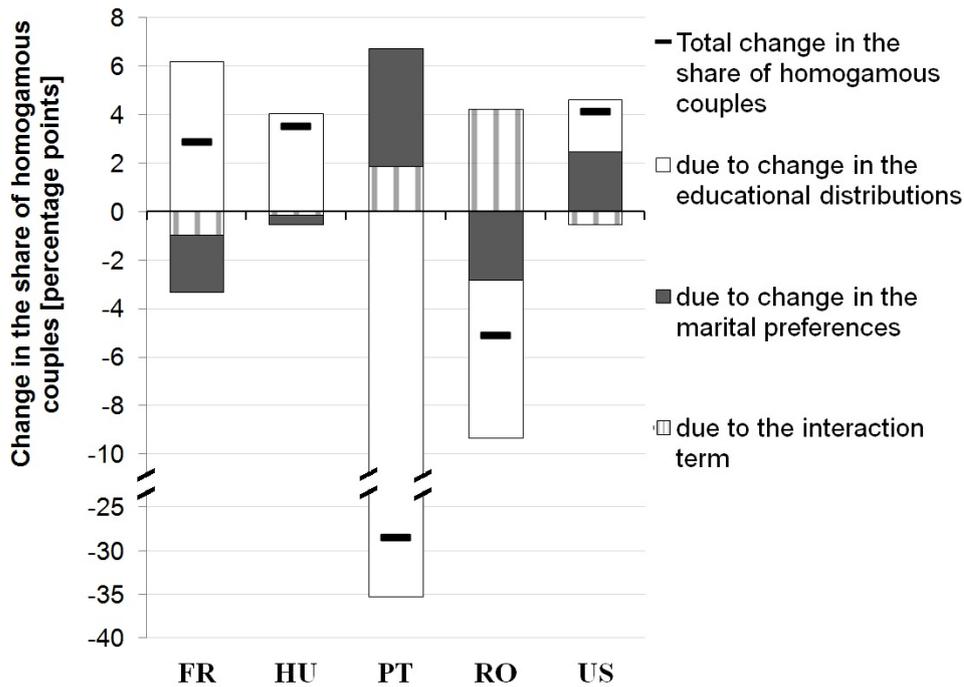

b) Long-horizon decompositions

*Notes:* A couple is considered to be educationally homogamous in any of the following three cases: (i) both of the spouses have at least a college degree, (ii) both of them completed high school, but neither of them obtained a tertiary level degree, (iii) neither of them completed high school. The short-horizon decomposition is based on the modified decomposition scheme presented by Eq. (3), while the components in the long-horizon decomposition are calculated as the sums of the components in the decade-specific short-horizon decompositions.

Figure 2. Decomposition of changes in the share of educationally homogamous couples between the late 1970s/early 1980s and the early 2010s - male partners are aged 25-39 years.

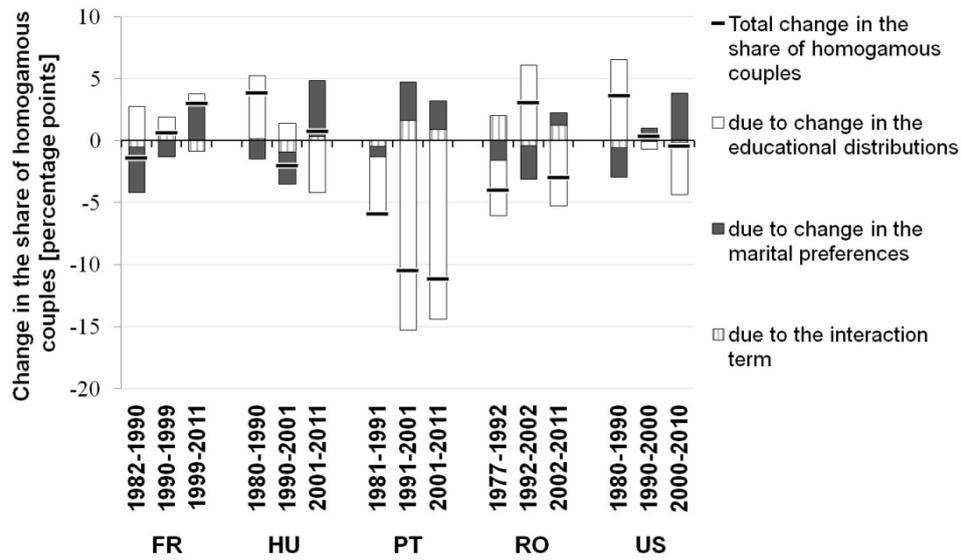

a) Short-horizon decompositions

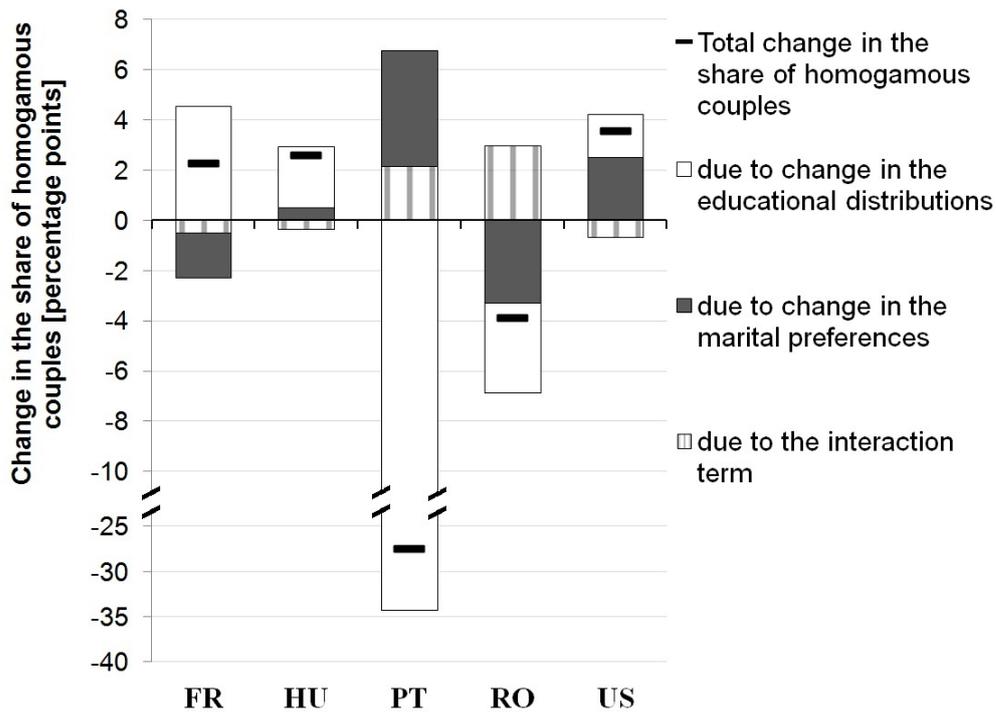

b) Long-horizon decompositions

Notes: Same as below Fig.1.